\documentstyle[aps,graphicx,floats,prl,amsmath]{revtex} 
\DeclareMathOperator{\trace}{Tr}

\begin{document}
\title{Trapping of Vibrational Energy in Crumpled Sheets}
\author{ Ajay Gopinathan, T.A. Witten }

\address{Dept. of Physics and James Franck Institute,
The University of Chicago,  Chicago, IL 60637}
\author{ S.C. Venkataramani}
\address{Dept. of Mathematics,
The University of Chicago,  Chicago, IL 60637}
\date{\today}

\maketitle

\begin{abstract}
 We investigate the propagation of transverse elastic waves in crumpled media. We set up the wave equation for transverse waves on a generic curved, strained surface via a Langrangian formalism and use this to study the scaling behaviour of the dispersion curves near the ridges and on the flat facets. This analysis suggests that ridges act as barriers to wave propagation and that modes in a certain frequency regime could be trapped in the facets. A simulation study of the wave propagation qualitatively supported our analysis and showed interesting effects of the ridges on wave propagation. 
\end{abstract}

\section{Introduction}
        Waves show distinctive behavior when they propagate through media that
are strongly heterogeneous.  Among these behaviors are localization \cite {soukoulis}, multiple
propagation paths and lensing \cite {blandford} .
        These phenomena depend strongly on the type of wave (electromagnetic,
Schroedinger, vibrational) and the type of heterogeneity.
       \par We treat a type of heterogeneity that arises naturally in materials,
but whose effect on waves has not been studied: crumpled membranes. 
        The disorder introduced by crumpling is much different from the random
disorder usually considered.  Instead of a structureless disorder, crumpled
sheets consist of uniform, flat facets bordered by strongly curved and
stretched borders called elastic ridges \cite{witten,witten2}.  Vibrational waves are expected to
propagate much differently in the facets and in the border regions.
        Crumpled sheets of newsprint paper are commonly used as an insulating and
protective enclosure, but we know of no characterization of their vibrational
isolation or transmissions properties. 
      \par  We anticipate various possible effects.  We imagine oscillating a point
on the membrane and investigating the response as a function of distance from
this source as compared with a flat, uniform sheet.  First, there are two
natural regimes of vibrational frequency to be considered.  Low frequencies are
expected to produce vibrations with wavelengths much longer than the facets or
ridges of the crumpled structure.  These long-wavelength waves might show
distinctive and subtle features, yet we shall not treat them in the present
paper.  These waves are not expected to depend heavily on the distinctive
facet-and-ridge structure of crumpled sheets.    
  \par      We shall consider instead the complementary regime of higher
frequencies having wavelengths smaller than the typical facets of the sheet. 
Several behaviors might be expected as compared to the uniform sheet.  On the
one hand, the elastic tension induced by the ridges might enhance the
propagation along the network of ridges.  On the other hand, the contrasting
nature of the ridges and the facets might serve as a barrier for propagation
out of or into a facet.  It is these phenomena that we address below.
    \par    In section II we briefly review classical thin membrane theory and gather the results we need. In section III we derive the wave equations for transverse waves on a generic curved, strained surface from a Langrangian formulation. Section IV is devoted to  studying the form of the dispersion relations in the two distinct regions (ridges and facets) and also the scaling behaviour of the dispersions with thickness of the sheet. This analysis first suggests the possibility of trapped energy in the facets with ridges serving as barriers. In section V we discuss in detail the simulations we performed and compare them to our analytical predictions. Finally we discuss the implications of our work and the scope for future work in section VI.

\section{Thin membrane theory}
Our findings are based on the classical theory of thin membranes \cite {landau}. If the thickness is small then we may assume that all variables may be integrated over the thickness of the membrane and it can be regarded as a two dimensional surface \cite {landau}. To define a surface uniquely up to overall translations and rotations it is sufficient to specify a metric tensor $g_{\alpha \beta}$ and a curvature tensor $C_{\alpha \beta}$ satisfying Gauss's {\it Theorema Egregium} and the Codazzi-Mainardi equations \cite{milman}. To see the physical significance of these tensors we first  note that the deviation of the metric tensor from the identity is simply the strain tensor
\begin {equation}
  g_{\alpha \beta} = \delta_{\alpha \beta} + \gamma_{\alpha \beta}
\end{equation}
The sum of the eigenvalues of the strain tensor is the (2D) expansion factor and their difference is the shear angle \cite{landau}. The eigenvalues of the curvature tensor are simply the inverses of the two principal radii of curvature of the surface. If we take a point $\vec{r} (x_1,x_2)$ where $x_1,x_2$ refer to the (2D) material co-ordinates of the point, the metric and curvature tensor are given by \cite{milman}
\begin {align}
  g_{\alpha \beta} & = (\partial_\alpha \vec{r}) \cdot (\partial_\beta \vec{r}) \\
  C_{\alpha \beta} & = \hat{n} \cdot (\partial_\alpha \partial_\beta \vec{r})
\end{align}
where $\partial_\alpha$ denotes differentiation with respect to the material co-ordinate $x_\alpha$, and $\hat{n}$ is the unit normal to the surface at that point. We now consider the forces and torques in the system necessary for mechanical equilibrium. They are given by the stress tensor $\sigma_{\alpha \beta}$ and the torque tensor $M_{\alpha \beta}$. These are related to the deformation tensors $\gamma_{\alpha \beta}$ and $C_{\alpha \beta}$, for sufficiently small strains by \cite{landau}
\begin {align}
  \sigma_{\alpha \beta} & = \frac{Yh}{1-\nu^2} [ \gamma_{\alpha \beta} + \nu \epsilon_{\alpha \rho}\epsilon_{\beta \tau}\gamma_{\rho \tau}] \\
  M_{\alpha \beta} & = \kappa [C_{\alpha \beta} + \nu \epsilon_{\alpha \rho}\epsilon_{\beta \tau} C_{\rho \tau}]
\end{align}
where $Y$ is the Young's modulus, $\nu$ is the Poisson ratio and $\kappa=Yh^3/(12(1-\nu^2))$ is the bending modulus of the elastic material. Here $h$ is the thickness of the sheet.  $\epsilon_{\alpha \beta}$ is the two dimensional antisymmetric tensor. Summation over repeated indices is implied. The conditions that guarantee that the strain and curvature tensors describe a physical surface in equilibrium are called the von Karman equations. The  von Karman equations \cite{karman} may be put in a form more suitable for our purposes as follows \cite{lobkovsky}

\begin {align}
  \det C_{\alpha \beta} & = \partial_\alpha \partial_\beta \gamma_{\alpha \beta} - \nabla^2 \trace \gamma_{\alpha \beta} \\
\partial_\alpha \partial_\beta M_{\alpha \beta} & =  \sigma_{\alpha \beta} C_{\alpha \beta}
\end{align}
The first equation is simply a restatement of Gauss' theorem which helps in making sure that the curvature and metric tensor describe a physical surface. The second von Karman (also called the stress von Karman) equation is a condition for physical equilibrium of the sheet.
Finally we need expressions for the elastic energy stored in the sheet. The stretching energy is given by \cite{lobkovsky}
\begin {equation}
E_s = \frac{1}{2} \int dxdy \, \sigma_{\alpha \beta} \gamma_{\alpha \beta}
\end{equation}
and the bending energy by 
\begin {equation}
E_b = \frac{1}{2} \int dxdy \,  M_{\alpha \beta} C_{\alpha \beta}
\end{equation}
Thus the total elastic energy of the sheet is given by
\begin {equation}
E_{tot} = \frac{1}{2} \int dxdy \, [ \sigma_{\alpha \beta} \gamma_{\alpha \beta} + M_{\alpha \beta} C_{\alpha \beta}]
\end{equation}

\section{The wave equation}

Let the equilibrium configuration of the sheet be given by $\vec{r}_0(x,y)$. This implies that $\vec{r}_0(x,y)$ satisfies the von Karman equations. To study tranverse waves we consider small displacements normal to the surface everywhere.
\begin {equation}
\vec{r}(x,y) = \vec{r}_0(x,y) + \epsilon u(x,y) \hat{n}
\end{equation}
where $u(x,y)$ is a smooth function of the material coordinates $x,y$. Considering only transverse waves is justified since it is known that in the thin limit ($h \ll 1$) the transverse and in-plane modes decouple \cite{pierce}. Our simulation results also indicate that for small perturbations the generated modes are {\it primarily} transverse and the fraction of energy in the in-plane modes is orders of magnitude smaller. The transverse perturbation $u(x,y)$ leads to a change in the total energy of the sheet of order $\epsilon^2$. The $\cal{O}(\epsilon)$ term should vanish because $\vec{r}_0(x,y)$ is the equilibrium configuration. Using equation (11) along with the defining equations (1)-(5) and equation (10) yields the $\mathcal{O}$$(\epsilon^2)$ change in energy
\begin{multline}
\delta E   =  \frac{1}{2} \int dxdy \, \left[  \kappa \left[ (u_{xx} - u (C_{xx}^0)^2)^2 +(u_{yy} - u (C_{yy}^0)^2)^2 \right] + 2 \kappa (1-\nu) u_{xy}^2 \right. \\
 + 2 \kappa \nu (u_{xx} - u (C_{xx}^0)^2) (u_{yy} - u (C_{yy}^0)^2) 
      + \frac{hY}{1-\nu^2} \left[ \gamma_{xx}^0 (u_x^2 + u^2(C_{xx}^0)^2 + \nu(u_y^2 + u^2(C_{yy}^0)^2)) \right. \\
   \left. \left.         + \gamma_{yy}^0 (u_y^2 + u^2(C_{yy}^0)^2 + \nu(u_x^2 + u^2(C_{xx}^0)^2)) 
       + u^2((C_{xx}^0)^2 + (C_{yy}^0)^2 + 2\nu C_{xx}^0C_{yy}^0) \right] \right]
\end{multline}
where $C_{xx}^0,C_{yy}^0,\gamma_{xx}^0,\gamma_{yy}^0$ are simply the respective components of the curvature and strain tensor in the equilibrium state. $u_\alpha$ is the derivative of $u$ with respect to the material coordinate $\alpha$. The $\cal{O}(\epsilon)$ term is simply the stress von Karman equation and hence vanishes as argued above. Cross terms involving  $C_{xy}^0$ and $\gamma_{xy}^0$ are absent because we have chosen our axes to be aligned with the principal axes of the curvature and strain tensor. Here we restrict ourselves to the case where the principal axes of the curvature and strain tensor are the same. This is true along the middle of the ridge \cite{witten2}. The general expression is more complex but this is sufficient for our purposes. We now consider $\delta E$ to be an effective potential for the field $u(x,y)$.
\begin {equation}
V = \delta E
\end{equation}
We can then define a kinetic energy term
\begin {equation}
T =  \frac{1}{2} \int dxdy \, \rho h ~ u_{t}^2
\end{equation}
where $\rho$ is the density of the material and $u_t \equiv \partial u/\partial t$. This leads to an effective Langrangian for the field given by
\begin {equation}
{\mathcal{L}}=  T - V
\end{equation}
The Euler-Lagrange equations for this system are then given by 
\begin {equation}
\frac{\partial \cal L}{\partial u} - \partial_{\mu} \frac{\partial \cal L}{\partial u_{\mu}} + \partial_{\mu}\partial_{\nu}\frac{\partial \cal L}{\partial u_{\mu \nu}} = 0
\end{equation}
The last term arises because of the presence of second derivatives of the field in the potential \cite{goldstein}. Using equations (12)-(15) in equation (16) gives us the local wave equation satisfied by the field $u(x,y)$.
\begin{equation}
-\rho h ~ u_{tt} = \kappa \nabla^4 u - A_x u_{xx} - A_y u_{yy} + B u
\end{equation}
where 
\begin{equation}
A_x = 2 \kappa (C_{xx}^0)^2 + 2 \kappa \nu (C_{yy}^0)^2 + \frac{Yh}{1-\nu^2} (\gamma_{xx}^0 + \nu \gamma_{yy}^0)
\end{equation}
and similarly for $A_y$. The coefficient of the last term is given by
\begin{multline}
B = \kappa (C_{xx}^0)^4 + \kappa (C_{yy}^0)^4 + 2 \kappa \nu (C_{xx}^0)^2 (C_{yy}^0)^2 
 + r [\gamma_{xx}^0 ((C_{xx}^0)^2 +  \nu (C_{yy}^0)^2) + \gamma_{yy}^0 ((C_{yy}^0)^2 +  \nu (C_{xx}^0)^2)] \\
 + r [ (C_{xx}^0)^2 + (C_{yy}^0)^2 + 2 \nu C_{xx}^0 C_{yy}^0] .
\end{multline}
where $ r \equiv Yh/(1-\nu^2)$. It is to be noted that all these coefficients are local, in that they depend on the curvature and strain at the point $(x,y)$ in the material co-ordinates. We now consider some limiting cases to understand the origin of the various terms in equation (17). First we consider the case when we simply have a flat sheet with no curvature or strain. Here  $ A_x,A_y$ and $B$ are identically zero, giving us
\begin{equation}
-\rho h ~ u_{tt} = \kappa \nabla^4 u 
\end{equation}
which correctly describes bending waves on a sheet with thickness $h$, bending modulus $\kappa$ and density $\rho$ \cite{landau}. Now we consider the case where we have a flat sheet with no curvature, $\kappa = 0$, and with some isotropic strain $\gamma$. Equation (17) then reduces to 
\begin{equation}
-\rho h ~ u_{tt} = \frac{Yh \gamma}{1-\nu} \nabla^2 u
\end{equation}
which describes a stretched membrane under tension proportional to $ Y\gamma$. The leading-order behaviour of the last term in equation (17) will be discussed in the next section.
\section{Dispersion relations}
A crumpled sheet consists of two distinct regions, the ridges and the flat facets that are bounded by ridges. The ridges are regions of large curvature and strain and most of the elastic energy of the sheet is stored here \cite{witten}. The flat facets on the other hand have small curvature and  strain from the ridges pulling on them \cite{witten2}. In this section we look at the dispersion relations for tranverse waves in the two regions.
\subsection{Ridges}
We take the material coordinate $x$ to lie along the width of the ridge and $y$ to lie along its length. We assume that the ridge is a region with uniform curvature and strain given by the values that they take at the midpoint. To obtain the dispersion relations we consider a plane wave, $ u \sim e^{i(kx - \omega t)}$, traversing the width of the ridge . Substituting this ansatz into equation (17) yields
\begin{equation}
 ak^4 + bk^2 + c = \omega^2 
\end{equation}
where 
\begin {align}
 a & = \frac{\kappa}{\rho h} \\
 b & = \frac{1}{\rho h} [ 2 \kappa C_{xx}^2 + 2 \kappa \nu C_{yy}^2 + r(\gamma_{xx} + \nu \gamma_{yy}) ] \\
 c & = \frac{1}{\rho h} [  \kappa C_{xx}^4 + r \gamma_{xx}C_{xx}^2 + r \nu \gamma_{yy} C_{xx}^2 + r C_{xx}^2  + \kappa C_{yy}^4 + 2 \kappa \nu  C_{xx}^2 C_{yy}^2 \nonumber\\
 & + r \nu \gamma_{xx}C_{yy}^2 + r  \gamma_{yy} C_{yy}^2 + r C_{yy}^2 + 2 r \nu  C_{xx} C_{yy}].
\end{align}
where $ r \equiv Yh/(1-\nu^2)$ as before. Considering a wave travelling in the perpendicular direction yields the same equation except with a $x$ and $y$ interchange in the coefficient $b$.
We now consider the leading order behaviour of the above coefficients in the limit of very small thickness , $h \ll 1$, where $h$ is measured in units of the ridge size. For this case the scaling of the midridge curvature and strain with thickness have been worked out by Witten and coworkers \cite{witten,witten2,lobkovsky}. We have the following results for 
\begin {align}
C_{xx} & \sim h^{-\frac{1}{3}} \\
C_{yy} & \ll C_{xx} \\
\gamma_{xx} & \sim  h^{\frac{2}{3}} \\
\gamma_{yy} & \sim h^{\frac{2}{3}}
\end{align}

Using these scaling laws along with the dispersion relation obtained above gives us the scaling behaviour of the dispersion curve
\begin{equation}
 a_0 h^2 k^4 + b_0 h^{\frac{2}{3}} k^2 + c_0 h^{-\frac{2}{3}} = \omega^2 
\end{equation}
where $a_0,b_0,c_0$ are positive real numbers independent of $h$. It is also to be noted that we would get the same form for the dispersion in the perpendicular direction of propagation. 
The explicit form of the leading order term in the coefficient $c$ is 
\begin{equation}
c = \frac{Y}{\rho (1 - \nu^2)} R^{-2}
\end{equation}
where $R = C_{xx}^{-1}$ is the transverse radius of curvature of the ridge. This is significant because the above expression is the square of the ``ringing frequency'' of a cylinder of radius $R$, which is the frequency of purely radial oscillations of the cylinder \cite{pierce}. Thus the last term in the dispersion represents the ``breathing mode'' of the ridge.
 Equation (30) tells us that, for large wavenumbers ($k \gg  h^{-\frac{2}{3}}$), the first term dominates leading to pure bending wave behaviour. This is reasonable, since for small wavelengths (or large $k$) the surface is locally flat and hence we expect behaviour like bending waves on a flat sheet. In the long wavelength regime ($k \ll  h^{-\frac{2}{3}}$), the constant term dominates leading to breathing modes of the ridge. This also prescribes a lower cutoff frequency for modes in the ridge. There is {\it no} regime where the second term  describing sound modes dominates. Thus the crossover leads directly from the bending wave regime to the breathing mode regime.
\subsection{Facets}
\begin{figure}
  \centering
  \includegraphics[width=3.2in]{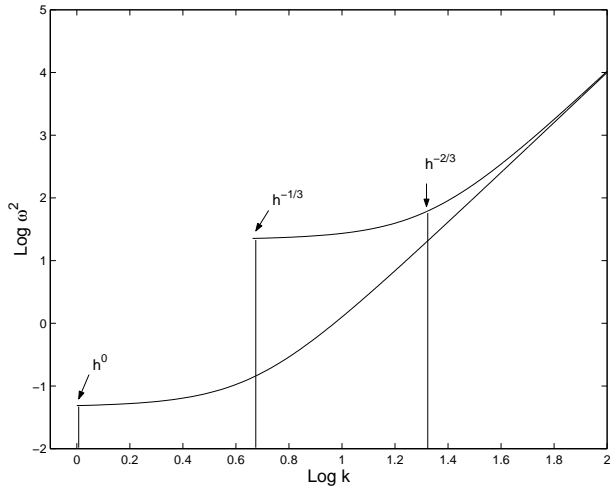}
  \vspace{1ex}
  \caption{Dispersion curves for the ridge (top curve) and facet (bottom curve) regions for $h=0.01$. Coefficients independent of $h$ ($a_0,b_0$ etc) have been set to 1. The lower $k$ cutoff for the ridge region is the inverse of the ridge width ($w^{-1} \sim h^{-1/3}$). The lower cutoff for the facet region is the inverse of the ridge length which we take as unity. Above $k \approx h^{-2/3}$ bending modes dominate in the ridge and modes in the facet can exist in the ridge too. }
  \label{fig:disp}
\end{figure}

The facets have some residual strain and curvature due to the ridges. Following the treatment of Lobkovsky {\it et al} \cite{witten2} we have the following scaling relations for the curvature and strain away from the ridge.
\begin {align}
C & \sim h^{\frac{1}{3}} \\
\gamma & \sim h^{\frac{4}{3}}
\end{align}
 Using these scaling laws, we follow the same procedure as in the last section to obtain the scaling form of the dispersion relation
\begin{equation}
 a_1 h^2 k^4 + b_1 h^{\frac{4}{3}} k^2 + c_1  h^{\frac{2}{3}} = \omega^2 
\end{equation}
Again we see two distinct regimes. For small wavelengths ($k \gg h^{-\frac{1}{3}}$), we have pure bending waves and for large wavelengths  ($k \ll h^{-\frac{1}{3}}$), we have breathing modes. Figure~\ref{fig:disp} shows the dispersion curves for both the ridges and facets. We notice that there are modes that exist on the facets whose frequencies are too low to propagate in the ridges. This prompts the question as to whether these modes can tunnel through the ridges to the other facets. 
\subsection{Skin Depth}
To address the question of whether the above mentioned modes are able to  penetrate the ridges we look at the skin depth of these modes in the ridge region. For a particular mode the frequency is the same in both regions ( because of matching at the interface ). If the frequency is less than the cutoff frequency in the ridge region, as is the case with the  modes we are looking at, then there will be an associated skin depth. This needs to be compared to the width of the ridge to see if these modes are trapped in the facet.
Consider a mode of frequency $\omega_0$. In the ridge region we have
\begin{equation}
 a_0 h^2 k^4 + b_0 h^{\frac{2}{3}} k^2 + c_0 h^{-\frac{2}{3}} = \omega_0^2 
\end{equation}
where $a_0,b_0,c_0$ are just real, positive numbers independent of $h$. We consider the case where  $\omega_0$ is just less than the cutoff frequency. Thus we take 
\begin{equation}
 \omega_0^2 = d_0 h^{-\frac{2}{3}}
\end{equation}
where $d_0 < c_0$. Using this we may now solve for $k$ in the  ridge region. The solution is
\begin{equation}
k = \big[ \frac{ -b_0 \pm (b_0^2 - 4 a_0 (c_0-d_0))^{\frac{1}{2}}}{2 a_0} \big]^{\frac{1}{2}} h^{-\frac{2}{3}}
\end{equation}
The requirement $d_0 < c_0$ ensures that $k$ has a nonzero imaginary part. Thus the imaginary part of $k$ scales as $h^{-\frac{2}{3}}$ and hence the skin depth
\begin{equation}
\zeta \sim [\mbox{Im} (k)]^{-1} \sim h^{\frac{2}{3}}
\end{equation}
The width of the ridge $w$, however scales as $h^{\frac{1}{3}}$ \cite{lobkovsky}. Thus we find that
\begin{equation}
\zeta \ll w
\end{equation}
This implies that the modes we considered ought to remain trapped in the facet for sufficiently thin sheets.
In the next section we examine the predicted scaling behaviour and the trapping phenomenon numerically.

\section{Simulations}
We model the elastic sheet as a triangular lattice of springs of unstretched length $a$ and spring constant $K$ following Seung and Nelson \cite{seung}. We incorporate bending rigidity by assigning an energy $J(1 - \hat{n_1} \cdot \hat{n_2})$ to every pair of adjacent lattice triangles with normals $\hat{n_1}$ and $\hat{n_2}$. Sueng and Nelson showed that when the strains are small and the radii of curvature are large compared to the lattice constant, this model membrane bends and stretches like an elastic sheet of thickness $h = a \sqrt{8J/K}$ with Young's modulus $Y = 2Ka/h\sqrt{3}$ and Poisson ratio $\nu=1/3$. The bending modulus is $\kappa = J\sqrt{3}/2$. This model has been extensively used to study the static scaling properties of ridges \cite{lobkovsky,witten2,brian}. We use it in this paper to study the dynamics of wave propagation by adding a unit mass to every lattice point. This models the kinetic term in our Lagrangian and assigns to the sheet, a mass per unit area.
\par 
To produce connected ridges and facets we construct a very simple shape, namely the regular tetrahedron, by joining the edges of a triangular sheet \cite{lobkovsky}. By using a conjugate gradient minimization routine we get the tetrahedron close to its equlibrium position. We then add in damping and a small perturbation (a tap on one of the facets) and then wait for the system to attain a state of equilibrium.
Once equilibrium is attained we now remove the damping and perturb the system ( tapping a few lattice points at the center of a facet by giving the three central particles of the facet equal initial velocities normal to the surface) so as to generate waves and study their propagation. To measure the energy being carried by the waves we first define an energy that measures the deviation from the equilibrium state.
\begin{align}
E_{dev} = & \sum \frac{1}{2} K (x_i - x_i^0)^2 +  \sum 2J ( \sin (\theta_k/2) -  \sin (\theta_k^0/2) )^2 \nonumber \\ &+ \sum \frac{1}{2} v_i^2
\end{align}
where $x_i$'s refer to the positions of the point unit masses, $\theta_i$'s are supplements of the angles between pairs of plaquets  and $v_i$'s are velocities of the masses. The superscript indicates the value of the variable at equilibrium. After an initial drop ($3-4\%$) this energy is conserved up to $1\%$ of its value and hence is a reasonably reliable indicator of the energy associated with deviations. 
\par
\begin{figure}
  \centering
  \includegraphics[width=3.2in]{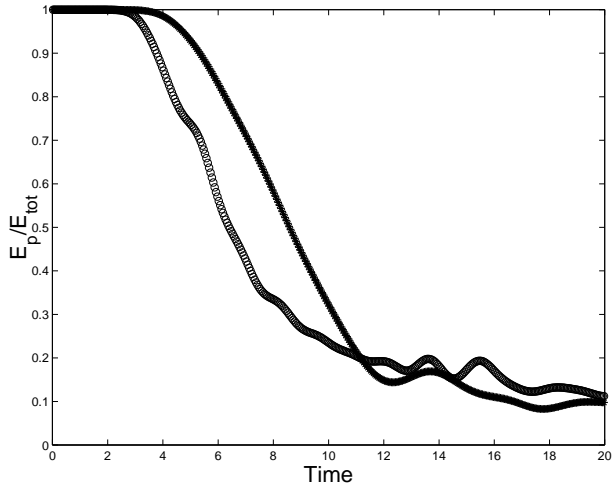}
  \vspace{1ex}
  \caption{Normalized energy of the perturbed facet as a function of time. $E_p$ stands for the deviation energy of the perturbed facet. $E_{tot}$ is the total deviation energy of the tetrahedron. Curves are for values of impulse 0.01 (+),0.001 (*) and 0.3 (o). Note that the first two are below the threshold (0.1) and the last one is above. }
  \label{fig:taps}
\end{figure}

 
For large initial perturbations we may also expect the initiation of in-plane waves. To avoid these waves we first estimate what initial impulse causes a displacement comparable to the lattice spacing, by assuming that the energy of the impulse is transferred to stretching and bending energy in the region of the tap. For typical values ($a=1$,$K=1$,$J=0.05$) we obtain the limiting value of the initial impulse to be $0.1$.  Figure~\ref{fig:taps} shows the normalized energy of the perturbed facet for different tap strengths both above and below the limiting value. The simulation was done with the above mentioned parameter values for a tetrahedron of size $10a$. We notice that the curves below the threshold lie on top of each other. Thus tapping below the threshold strength gives us robust, characteristic curves that are independent of the initial impulse. From here on we consider only simulations where we tap below the threshold. Looking at the stretching and bending energies in this regime also tells us that stretching contribution is small and hence the modes are mostly transverse. We also check this fact by looking explicitly at the particle motion.

Now we consider the time dependence of the energy of the perturbed facet. Typical curves are like those in figure~\ref{fig:taps} with initial impulse below the threshold. For an initial interval the energy remains close to 1. It then begins to drop off and we see a straight line segment  to the curve. The initial interval simply indicates the time it takes the fastest mode to reach the edge of the facet. The straight line segment on the other hand is presumably where all modes are contributing giving us a constant energy flux out of the facet and hence a constant slope in the energy versus time graph.

\begin{figure}
  \centering
  \includegraphics[width=3.2in]{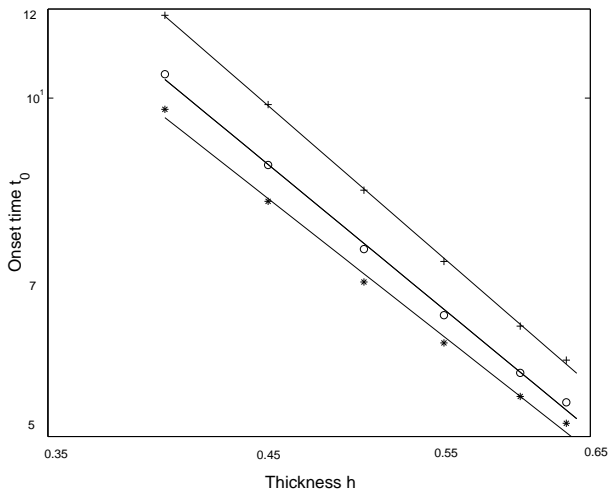}
  \vspace{1ex}
  \caption{Log-log plot of onset times as a function of thickness for different $p$. Circles ($p=0.9$), Pluses ($p=0.85$), Asterisks ($p=0.92$). The lines are best fit to the data points. Slopes of the lines (top to bottom) are -1.55,-1.48 and -1.41 respectively.}
  \label{fig:onset}
\end{figure}

To measure these two quantities we first define an onset time $t_0$ where the energy begins leaking out to the other facets. We take this as the time for the energy of the perturbed facet to drop to a fraction $p \sim 0.9$. We choose 0.9 because it is close enough to 1 so as to justify the name onset time but still far enough from 1 to get a good spread of data.  We now wish to study how this onset time scales with the thickness of the sheet. We obtain data for different thicknesses by running the simulation with the parameter $a$ kept fixed and changing $J,K$ such that the Young's modulus remains constant while the bending modulus obeys $\kappa \sim h^3$. Note that since the unit masses are not changed the density per unit area ($\rho h$) remains constant. The onset time should be proportional to the inverse of the largest group velocity. From  figure~\ref{fig:disp} we see that the fastest modes should be predominantly bending modes. Hence we expect
\begin{equation}
t_0 \sim [\frac{d\omega}{dk}]^{-1} \sim (\sqrt{\kappa/\rho h})^{-1} \sim h^{-3/2}
\end{equation}
Figure~\ref{fig:onset} shows a plot of the onset times for various thicknesses and for different values of $p$. Best fit lines for slightly different values of $p$ are fairly parallel and hence the exact choice of $p$ does not affect the power law crucially. From the graph we deduce a power law of $t_0 \sim h^{-1.48 \pm 0.07}$ which agrees very well with the theory. The error quoted is the variation of the slope with $p$.

\begin{figure}
  \centering
  \includegraphics[width=3.2in]{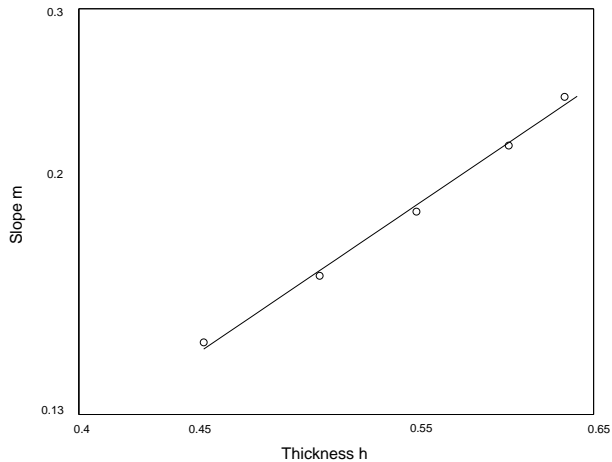}
  \vspace{1ex}
  \caption{Log-log plot of slope $m$ as a function of thickness. The line is a best fit to the data points.}
  \label{fig:slope}
\end{figure}

The scaling derived above does not reflect the presence of ridges. The onset time is a characteristic of the velocity of propagation in the facet and we would get an identical scaling law for a flat sheet. To identify the effects of the ridge we consider the portion of the curve that shows the energy leaving the facet. As mentioned before there is a straight line segment to this curve where we may assume all modes are contributing. We denote the slope of this line segment as $m$ and study how this quantity varies with thickness.
We also expect  that if the slopes ($m$) are dominated by the bending wave contribution then
\begin{equation}
m \sim [\frac{d\omega}{dk}] \sim \sqrt{\kappa/\rho h} \sim h^{3/2}
\end{equation}
Figure~\ref{fig:slope} shows the expected behaviour with a power $h^{1.46 \pm 0.1}$. Again this is exactly what we expect for a flat sheet. This could imply that the ridges do not affect any of the characteristics of wave propagation. 

\begin{figure}
  \centering
  \includegraphics[width=3.2in]{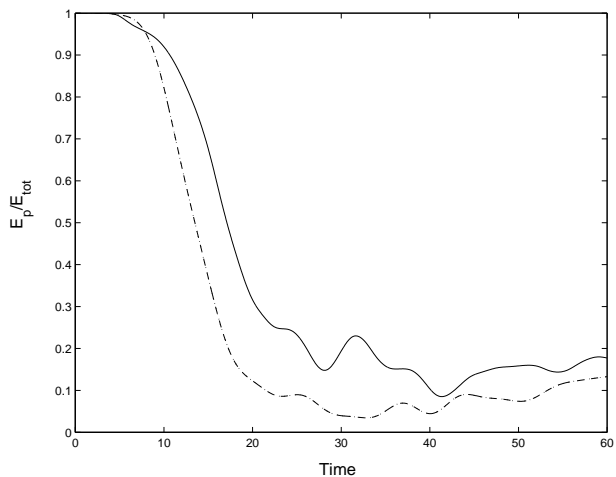}
  \vspace{1ex}
  \caption{Normalized energy of the perturbed facet as a function of time. $E_p$ stands for the deviation energy of the perturbed facet. $E_{tot}$ is the total deviation energy. Solid line - tetrahedron. Dashed line - flat sheet.}
  \label{fig:falltfl}
\end{figure}

To see if this really is the case we explicitly compare curves obtained for a tetrahedron and a flat sheet under appropriate strain in figure~\ref{fig:falltfl}. The curve for the flat sheet is obtained by using a flat triangular sheet obtained by unfolding the tetrahedron. The sheet is thus divided into four equal ``facets'' with the central one being perturbed. The sheet is put under strain by stretching the edges slightly and then clamping them. The magnitude of strain is set to be equal to that in the facets of the tetrahedron. The curves show the normalized energy of the same perturbed facet in the tetrahedral and flat geometry. From figure~\ref{fig:falltfl} we can draw one main conclusion. The rate at which energy leaves the facet seems appreciably less in the case of the tetrahedron. 

\begin{figure}
  \centering
  \includegraphics[width=3.2in]{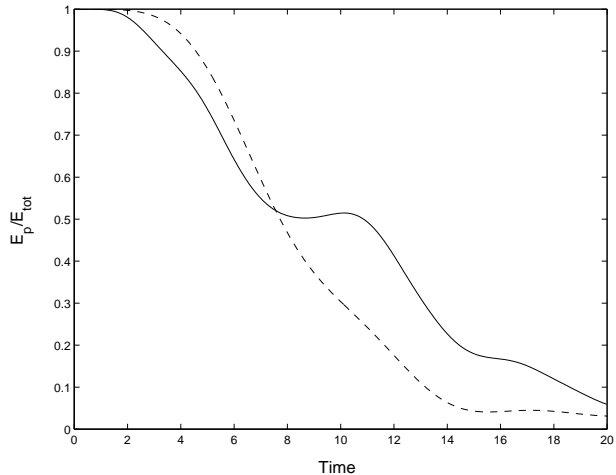}
  \vspace{1ex}
  \caption{Normalized energy of a section of the perturbed facet as a function of time. $E_p$ stands for the deviation energy of the subsection of the perturbed facet. $E_{tot}$ is the total deviation energy. Solid line - tetrahedron. Dashed line - flat sheet.}
  \label{fig:falltflsm}
\end{figure}

We can imagine three possible scenarios which inhibit the flow of energy from the perturbed facet. Firstly, significant reflections may occur at the ridge thus confining energy to the perturbed facet. Secondly, the energy may preferentially propagate through the centers of the ridges where the curvature is least, i.e. the large curvatures close to the vertices may inhibit propagation. Thirdly the energy may simply propagate preferentially along the ridge and not into the neighbouring facet. The first two mechanisms follow qualitatively from our analytical treatment while the third would invalidate our analysis. To check for the possibility of reflections we look at the energy of a smaller triangular section within our perturbed facet. Figure~\ref{fig:falltflsm} shows the comparison between the tetrahedral case and the flat sheet case. We immediately notice a bump in the curve for the tetrahedron indicating that energy flows back in, suggesting the occurence of reflections.

\begin{figure}
  \centering
  \includegraphics{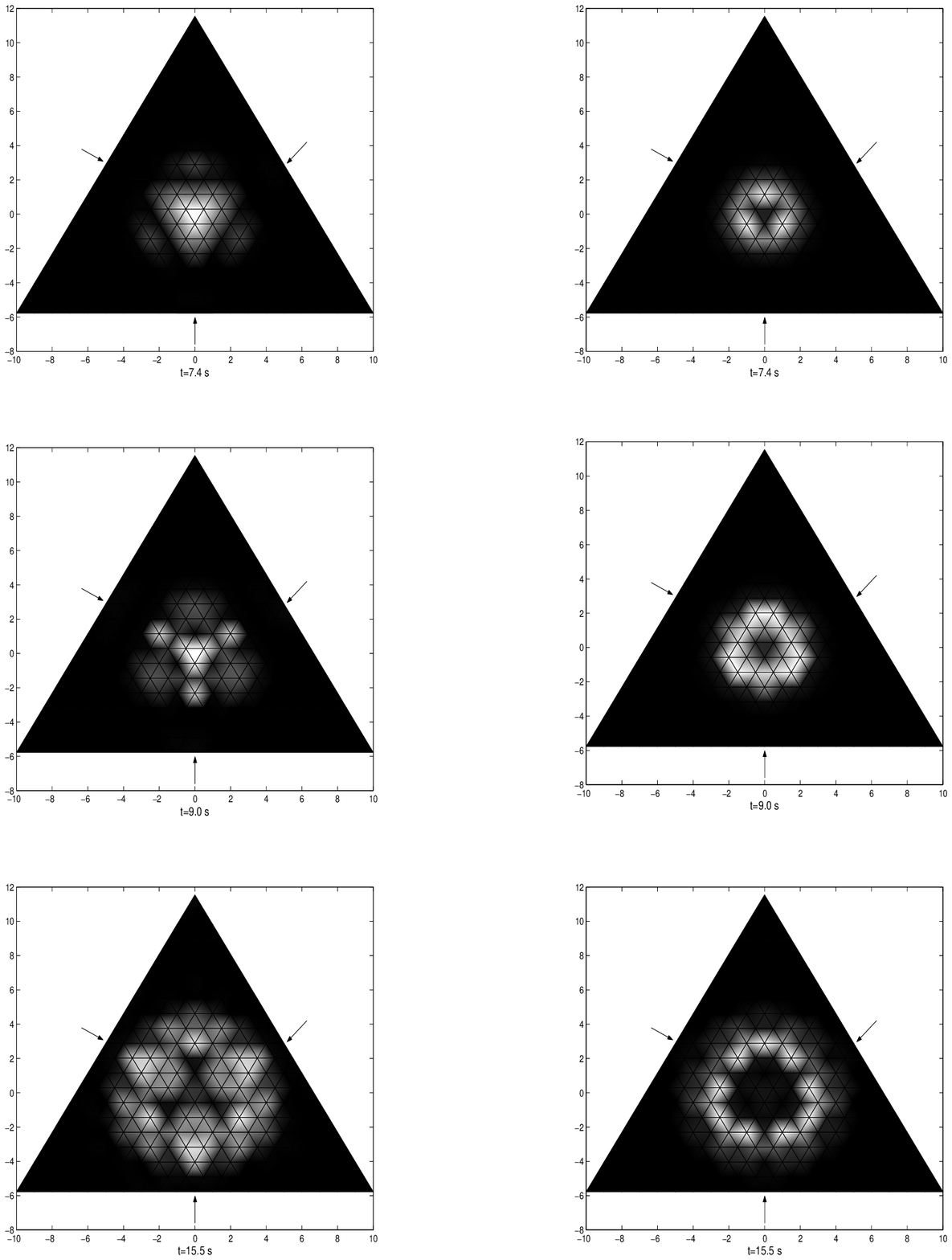}
  \vspace{1ex}
  \caption{Time sequence of energy density profiles for the tetrahedron (left) and flat sheet (right). For purposes of display the lattice points of the tetrahedron have been mapped to their undeformed triangular state. Arrows mark the vertices of the perturbed facet of the tetrahedron and the same region on the flat sheet for comparison. The brightest regions in each snapshot represent the regions of highest energy density in {\it that} snapshot.}
  \label{fig:p2}
\end{figure}

However to note the presence or absence of other features we looked explicitly at the spatial energy distribution as a function of time. Figure (\ref{fig:p2}) shows a sequence of snapshots of the spatial energy distribution for the tetrahedron on the left and a flat sheet on the right. The tetrahedron has been folded out and hence the central triangular face, marked at the vertices by arrows, is the perturbed face. Ridges form the sides of this face. Arrows in the flat sheet sequence simply help to identify the same region. Note that the initial excitation is confined to the central three particles. 
We see that in the flat sheet case, the energy simply spreads outwards. In the tetrahedral case though, we notice reflections ($t=6.5$ to $t=9$) and also the preferential propagation of energy through the middle of the ridges ($t=15.5$ ). Thus we see the qualitative effects predicted by our analysis. However we never see the complete trapping of energy. This could be due to the limited size of the simulation as well as the difficulty in producing wavelengths in the right regime. 

\begin{figure}
  \centering
  \includegraphics{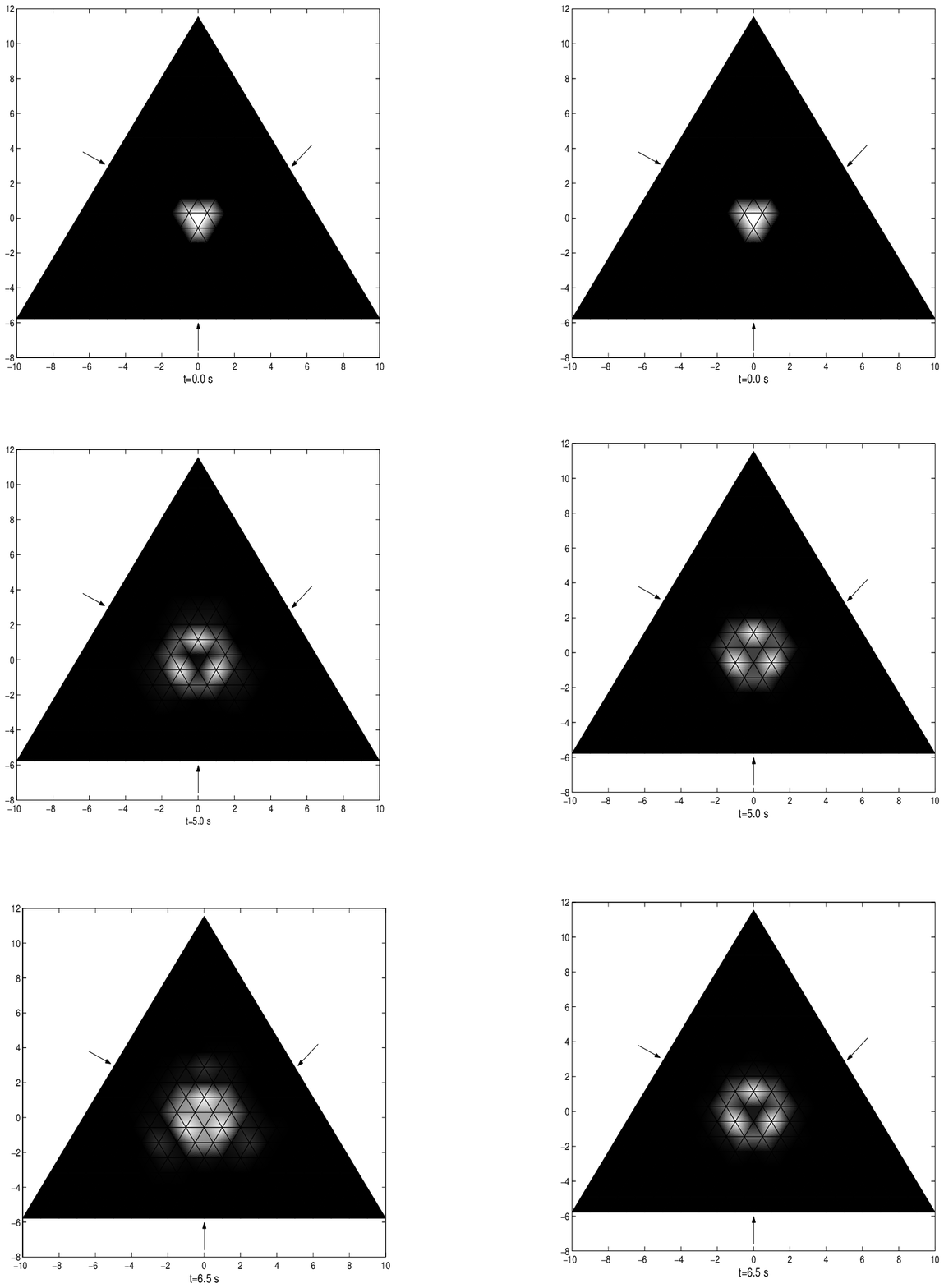}
  \vspace{1ex}
  \label{fig:p1}
\end{figure}

\section{Conclusion}
 In this paper we have taken the first step toward  understanding  the dynamics of crumpled elastic sheets. We find that the ridge structures in crumpled sheets qualitatively modify the propagation of transverse waves. Despite the stress induced by the ridges, our analysis indicates that the stretched string or drumhead modes normally associated with a stretched membrane are not important in a crumpled sheet. Our analysis suggested that the ridges would act as barriers to propagation and also that there could be modes that are trapped in the facets. These modes are unable to penetrate the ridges because the associated skin depth is much less than the ridge width in the thin limit. Our numerical simulations gave us qualitative evidence that the above picture was right. Future work could focus on subjecting our predictions to a more thorough simulation analysis which could support our findings in a more quantitative fashion as well as identify the trapping of modes in the right frequency regime. It would also be interesting to extend the analysis to cover the low frequency regime.

\section{Acknowledgements}
 The authors would like to thank Denis Chigirev and Vijay Patel for developing early versions of the simulation program used. AG would like to thank members of the Witten group and Brian DiDonna in particular for enlightening conversations. This work was supported in part by the National Science Foundation under
Award number DMR-9975533 and in part by the National Science Foundation's MRSEC
Program under Award Number DMR-980859

\end{document}